# Time, what is it? Dynamical Properties of Time[1]

Oleinik V.P.[*], Borimsky Yu.C.[*], Arepjev Yu.D.[**]

[*]Department of General and Theoretical Physics,
National Technical University of Ukraine "Kiev Polytechnic Institute",
Prospect Pobedy 37, Kiev, 03056, Ukraine
[**]Institute of Semiconductor Physics, National Academy of Sciences,
Prospect Nauky 45, Kiev, 03028, Ukraine; e-mail: yuri@arepjev.relc.com
Received January 10, 2000

**Abstract**. The phenomenon of local dynamical inhomogeneity of time is predicted, which implies that the course of time along the trajectory of motion of a particle in the inertial reference frames moving relative to each other depends on the state of motion of the particle under the influence of a force field. As is seen from the results obtained, the ability to influence the course of time represents one of the most fundamental properties of any material system intrinsically inherent in it by the very nature of things, which manifests itself when the system interacts with force fields. The inferences of the paper are not based on the use of any hypotheses and strictly follow from relativistic equations of motion. The dependence of the course of time upon the behaviour of physical system is, thus, a direct consequence of causality principle, relativity principle and the pseudoeuclidity of space-time. The results obtained confirm the Kozyrev hypothesis that time has physical properties and open up radically new opportunities for the efficient control of physical processes. It is demonstrated with point particle that the change in the course of time results in the appearance of an additional force acting on the particle. A general conclusion is drawn on the basis of the theory advanced that relativistic equations of motion for any kind of matter contain information about the physical properties of time which are, thus, of dynamical nature.

## 1. Introduction

Though all the events and processes in the world happen in space and in time and, hence, the laws that govern space-time connections are the most general and hold for all the forms of matter, time still remains one of the most mysterious concepts of physics, the essence of which is not adequately revealed up till now [1-4].

It is commonly supposed that time defines only the duration of physical processes and characterizes the consecutive order in which one event changes the other. Time is absolutely passive; it exists independently of material processes. At the same time, the run of all the events and processes, occurring in the world, is subject to it.

New possibilities for clearing up the physical gist of time are afforded by the special theory of relativity (STR), which combines space and time into a single whole – the 4-dimensional Minkowski space. Time and space coordinates are considered in STR as the equal in rights and self-dependent quantities defining the position of elementary events in space-time. Time, on the other hand, stands out in relation to space coordinates. The special role of time is due, from the viewpoint of geometry, to the pseudoeuclidity of geometry of the 4-dimensional space. From the physical point of view, it is associated with the dynamical principle (the causality principle), according to which the state of motion of a physical system at an instant of time $t$ unambiguously defines its behaviour at the following instant of time $t+0$. The significance of the dynamical principle lies in the fact that it relates the temporal evolution of system to the physical processes caused by force fields and in doing so it allows one to determine the course of time in the system, its possible dependence upon the character of physical processes, and not just the sequence of events and their duration.

In connection with the dynamical principle the question arises as to whether there exist the physical properties of time, i.e. the question of whether the physical processes taking place in a system can influence the course of time in it. The existence of the physical properties of time is supported by the following simple considerations [5]. The force fields (gravitational, electromagnetic etc.), created by material bodies in surrounding space, change the space endowing it with physical properties. Because space and time are indissolubly related to each other to form a single whole, the presence of a force field in space must necessarily result in the appearance of physical properties of time caused by the motion of a body in this field. This conclusion follows, in particular, from the Lorentz transformations, in accordance with which the temporal coordinate of some event in one inertial reference frame is expressed in terms of the temporal and spatial coordinates of this event in the other. Such an entanglement of coordinates indicates that time, as well as space, becomes a direct participant of physical processes.

The idea about the existence of the physical properties of time belongs to Kozyrev N.A [6]. According to the results of theoretical and experimental investigations, conducted by Kozyrev and his followers [6-9], events can occur not only in time, but also with the aid of time; in this case information is transmitted not via force fields, but through the temporal channel, the transfer of information occurring instantaneously. The appearance of extra forces, associated with the physical properties of time and capable to produce work, indicates that time can serve as an energy source. Let us point to papers [8,9], in which the physical properties of the world of events are discussed in detail and the problem of direct experimental research of the physical properties of time is formulated with the aim to ascertain the relations of a new type between phenomena and to discover new methods of changing the state of substance. The hypothesis on the existence of physical properties of time and its consequences are considered from a qualitative point of view in the paper by A.M. Bych [10].

For the purpose of constructing the rigorous and consistent theory taking into account the possible appearance of the physical properties of time, one should turn to dynamics. As was noted above, it is the dynamical principle that relates the evolution of a system in time to the action of the force fields. As A.A. Logunov underlines, "if for some form of matter we have the laws of its motion in the form of differential equations, then these equations contain information on the structure of space and time" [4]. Apparently, the dynamical equations should contain information not only about geometrical, but also about physical properties of space-time [5,11].

It is shown in this work on the basis of relativity principle and relativistic equations of motion of classical point particle that, when a particle moves in an external electromagnetic field, there arises the phenomenon of local dynamical inhomogeneity of time, having relative character. The essence of this phenomenon is that the course of time along the trajectory of motion of particle depends on the character of its motion. The inferences of the paper are illustrated by the example of homogenous electric and magnetic fields. It is noted that the phenomenon of time inhomogeneity does not result in violation of the total energy conservation law of the particle when it moves in arbitrary homogeneous external field. The results of the paper indicate that the physical properties of time are not preset a priori, they are inevitably created by material systems owing to their movement under the influence of the force fields.

The existence of the phenomenon of the dynamical inhomogeneity of time confirms Kozyrev's idea that time has physical properties, and opens the way to the quantitative explanation of the results of experimental investigations [6-8] without resort to any additional hypotheses.

The basic ideas of this paper are briefly outlined in [5,12,13].

## 2. Local Dynamical Inhomogeneity of Time

Let us consider the motion of a point particle from the viewpoints of two inertial reference frames $K$ and $K'$ moving relative to each other. Galilean coordinates are put into correspondence to each of them and it is assumed that the $y, z$ - axes and the $y', z'$ -axes are parallel to each other, the $x$ -axis coincides with the $x'$ -axis, the frame of reference $K'$ moving relative to the reference frame $K$ with a velocity $v_0$ along the $x$ - axis and coinciding with the $K$ at the instant of time $t = t' = 0$. The Lorentz transformations for coordinates are of the form:

$$t = \gamma(t' + v_0 x'), \qquad x = \gamma(x' + v_0 t'), \qquad y = y', \qquad z = z', \qquad \gamma = (1 - v_0^2)^{-1/2}. \qquad (1)$$

If in the empty space the motion of merely one particle is considered, then there are no scales of time and distances other than the time and space intervals connected with the motion of the same particle relative to some frame of reference. For this reason, it is these quantities which are adopted as the time and distance scales. Denote by $dr$ the increment of the position vector of particle in the frame of reference $K$ for the time $dt$, and by $dr'$ and $dt'$ the analogous quantities relating to the motion of particle in the reference frame $K'$. Then the quantities $\frac{dr}{dt} \equiv u(t)$ and $\frac{dr'}{dt'} \equiv u'(t')$ will have the meaning of the particle velocities in the reference frames $K$ and $K'$, respectively. Calculating differentials of the left and right parts of the former of the relationships (1), we obtain:

$$dt = \gamma(1 + v_0 u'_x(t')) dt'. \qquad (2)$$

From the inverse Lorentz transformations we can derive similarly:

$$dt' = \gamma(1 - v_0 u_x(t)) dt. \qquad (3)$$

The quantities $dt$ and $dt'$ in (2) and (3) have the following physical sense: they are time intervals between two infinitely close events associated with the particle motion in a path and considered in the inertial reference frames $K$ and $K'$, respectively. The following equality results from comparison of (2) and (3):

$$\gamma^2 (1 + v_0 u'_x(t'))(1 - v_0 u_x(t)) = 1, \qquad (4)$$

which leads to the velocity addition rule:

$$u_x(t) = \frac{u'_x(t') + v_0}{1 + u'_x(t') v_0}. \qquad (5)$$

Consider the case when the particle is at rest in the frame $K'$, i.e. $u'(t') = 0$. Then the quantity $dt'$ coincides with the proper time $d\tau$ of the particle, and equality (2) can be written as

$$d\tau = \sqrt{1 - v_0^2}\ dt. \qquad (6)$$

Relationship (6) describes the slowing down of the proper time $d\tau$ of the particle as compared to the time $dt$ measured by the $K$-observer moving with a velocity $v_0$ relative to the particle. The phenomenon described by (6) is characterized by the following features: firstly, the slowing down of the course of proper time is a purely kinematic effect and its magnitude depends only on the relative velocity of the reference frames $K$ and $K'$ and, secondly, this phenomenon is of a global character: the course of time in the reference frame $K$ at all points of space is changed to the same magnitude in comparison with the proper time.

Formula (6) refers to the case when the particle moves with a velocity $v_0 = const$ relative to some inertial frame of reference. The quantities $d\tau$ and $dt$ in (6) represent the time intervals measured in the inertial frames: $d\tau$ in the frame of reference, in which the particle is at rest, and $dt$ in the reference frame, relative to which the particle moves with velocity $v_0$. It should be emphasized that the phenomenon described by (6) is of a global character by virtue of special choice of inertial reference frames, which results in their relative velocity coinciding with the velocity of the particle.

Let now there be given a homogeneous magnetic field

$$\boldsymbol{B}' = (0,\ 0,\ B') = const. \tag{7}$$

in an inertial frame of reference $K'$. Solution to the relativistic equations of motion for a point particle with charge $e$, moving in the plane $z' = 0$, may be represented as (see [14])

$$u'_x = u'_0 \cos\varphi', \quad u'_y = -u'_0 \sin\boldsymbol{j}', \tag{8}$$

where $u'_0 = const$, $\omega' = eB'/m$, $\varphi' = \omega't' + \alpha$, $\alpha = const$, $m$ is the relativistic mass of particle. Write down the position vector of the particle in the form

$$\boldsymbol{r}'(t') = \boldsymbol{r}'_0 + \frac{u'_0}{\omega'}(\sin\varphi', \cos\varphi', 0), \tag{9}$$

$\boldsymbol{r}'_0 = const$. According to the formulas above, in the reference frame $K'$ the particle moves round a circle with constant velocity: $|\boldsymbol{u}'(t')| = u'_0$. Since to the arcs of the same length $ds'$ which the particle goes in the frame $K'$ in magnetic field (7) there correspond the time intervals $dt'$ ($dt' = ds'/u'_0$) equal in magnitude, the $K'$-observer can use the physical process under study for making ideal clock. The uniform course of this clock is due to the fact that all the instants of time on the time axis are physically equivalent.

However, the course of this clock proves to be irregular from the point of view of the reference frame $K$. Really, it is seen from (2) and (8) that

$$dt = g(t')dt', \qquad g(t') = \gamma(1 + v_0 u'_0 \cos\varphi'), \tag{10}$$

i.e. to the time intervals of the same length $dt'$ measured in the frame $K'$ there correspond the time intervals $dt$ in the frame $K$ depending on the choice of the instant of time ($dt = dt(t')$). Consider the events 1 and 2 related to the motion of two different particles in the magnetic field and described in the frame $K'$ by coordinates $(t', \boldsymbol{r}'_1(t'))$ and $(t', \boldsymbol{r}'_2(t'))$, respectively, where

$$\boldsymbol{r}'_n(t') = \boldsymbol{r}'_0 + \frac{u'_0}{\omega'}(\sin\varphi'_n, \cos\varphi'_n, 0), \qquad \varphi'_n = \omega't' + \alpha_n, \quad (n = 1,2). \tag{11}$$

Let us calculate the difference of time coordinates of these events in the view of the $K$-observer:

$$t_2 - t_1 = 2\gamma v_0 \frac{u'_0}{\omega'} \sin\frac{\alpha_2 - \alpha_1}{2} \cos\left(\omega't' + \frac{\alpha_2 + \alpha_1}{2}\right). \tag{12}$$

As is seen from (12), the events 1 and 2 simultaneous in the frame of reference $K'$ cease to be simultaneous in the frame $K$. What is more, because the course of time in the view of the $K$-observer is not uniform, the quantity $t_2 - t_1$ oscillates with time at the rotation frequency of the particle in magnetic field, i.e. from the viewpoint of the $K$-observer event 1 takes place alternately before and after event 2.

Calculate the period $T$ of particle's motion from the viewpoint of the $K$-observer:

$$T = \int_{t'_0}^{T'+t'_0} g(t')dt' = \boldsymbol{g}T'. \tag{13}$$

Here $T' = 2\pi/\omega'$ is the period of motion of the particle in the frame $K'$, $t'_0$ is an arbitrary instant of time. Determine now the time interval $\Delta t$ in the reference frame $K$, which corresponds to the half of the period of motion of the particle in the view of the $K'$-observer:

$$\Delta t = \int_{t'_0}^{\frac{T'}{2}+t'_0} g(t')dt' = \gamma \frac{T'}{2}\left(1 - \frac{2}{\pi} v_0 u'_0 \sin(\omega' t'_0 + \alpha)\right). \tag{14}$$

As is seen from (14), $\Delta t \neq T/2$, the quantity $\Delta t$ depending on the choice of zero reading of time ($\Delta t = \Delta t(t'_0)$) and the difference $\Delta t - T/2$ being an oscillating function of $t'_0$. Thus, the instants of time cease to be physically equivalent from the viewpoint of the $K$-observer.

If in an inertial frame of reference $K'$ there is a homogeneous electric field $\boldsymbol{E'} = const$, then the solution to the relativistic equation of motion $\frac{d\boldsymbol{p'}}{dt} = e\boldsymbol{E'}$, obeying the condition $\boldsymbol{p'}\big|_{t'=0} = 0$, may be written as $\boldsymbol{p'} = e\boldsymbol{E'}t'$. Accordingly, the velocity of particle is expressed by

$$\boldsymbol{u'}(t') = \frac{\boldsymbol{a}t'}{\sqrt{1+a^2 t'^2}}, \qquad \boldsymbol{a} = \frac{e\boldsymbol{E'}}{m_0}, \tag{15}$$

$m_0$ being the rest mass of particle. On the strength of Lorentz transformations we obtain:

$$dt = \gamma\left(1 + \frac{v_0 a_x t'}{\sqrt{1+a^2 t'^2}}\right)dt'. \tag{16}$$

From here it is seen, that $dt > \gamma dt'$ at $a_x t' > 0$ and $dt < \gamma dt'$ at $a_x t' < 0$, i.e. when the particle moves in a homogeneous electric field, the course of time depends on the direction of projection of the electric field strength on the direction of relative movement of reference frames. For comparison, let us calculate the differential of proper time of particle when it moves in the electric field:

$$d\tau = \sqrt{1 - (\boldsymbol{u'}(t'))^2}\, dt' = dt'/\sqrt{1+a^2 t'^2}. \tag{17}$$

According to the last formula, $d\tau < dt'$, with $d\tau/dt' \to 0$ at $t' \to \infty$.

As is seen from the examples considered above of motion of a charged particle in an external electromagnetic field in inertial reference frames moving relative to each other, the purely kinematic change in the course of time, dependent only on the speed of relative movement of reference frames, is accompanied by a dynamical one, which depends on the character of motion of the particle. The peculiarities of this phenomenon (we call it the effect of dynamical inhomogeneity of time) may be revealed in the general case, for arbitrary motion of particle, starting from relationships (2) and (3). According to (2), the course of time in the frame of reference $K$ depends not only on the velocity of relative movement of the reference frames, but also on the state of motion of the particle in the frame $K'$. Let for definiteness $v_0 > 0$. Then the time in the frame $K$ is running faster than in the frame $K'$ at $u'_x > 0$ ($dt > \gamma dt'$), and slower at $u'_x < 0$ ($dt < \gamma dt'$). The course of time in the frames of reference $K$ and $K'$ is identical (i.e. $dt = dt'$) provided $u'_x = \left(\sqrt{1-v_0^2} - 1\right)/v_0 \equiv u^*$ (or $u_x = -u^*$, see the velocity addition rule (5)). The time in the frame $K$ is running faster than in the frame $K'$ at $u'_x > u^*$ (or at $u_x > -u^*$) and slower at $u'_x < u^*$ (or at $u_x < -u^*$). It should be emphasized that here we deal with the change in the course of time in one inertial frame of reference as compared to the other not in the whole space at once, but locally – at the point of the particle localization. The dynamical inhomogeneity of time is, thus, of a local character: it manifests itself only along the trajectory of motion of particle. Generalizing the result obtained to a system of arbitrary number of particles, one may assert that in the regions of space filled by particles the course of time essentially depends upon the behaviour of particles.

According to the results given above, the dynamical inhomogeneity of time follows immediately from the Lorentz transformations and equations of motion. If the course of time depends on the state of motion of particles, then, apparently, there should take place a back influence of the time distorted by physical processes on the character of these processes.

Note that, according to (1),

$$\frac{dx}{dx'} = \gamma \left( 1 + \frac{v_0}{u'_x(t')} \right). \tag{18}$$

For a classical particle, on the basis of the Galilean transformations, we can derive:

$$dx = u_x \, dt = (u'_x + v_0) \, dt, \qquad dx' = u'_x \, dt.$$

From here it follows that $dx = dx' + v_0 \, dt$ and

$$\frac{dx}{dx'} = 1 + \frac{v_0}{u'_0}.$$

Up to the factor $\gamma$ the latter equality coincides with (18).

To clarify the physical gist of the phenomenon of dynamical inhomogeneity of time and its dissimilarity from the effect of slowing down the proper time of particle, let us consider again a particle moving arbitrarily relative to an inertial frame of reference $K$. We put into correspondence to this motion the time-space interval $ds^2 = dt^2 - d\mathbf{r}^2$, where the quantities $d\mathbf{r}$ and $dt$ are assumed to be, as at the beginning of this section, the increment of the radius-vector of the particle and the time interval, corresponding to it, measured in the reference frame $K$. Then the quantity $ds^2$ can be written as $ds^2 = dt^2 \left(1 - \mathbf{u}^2(t)\right)$. Denote by $S$ the reference frame, in which the particle under study is at rest. In this frame of reference, which is noninertial, the following equality holds: $ds^2 = d\tau^2$, with $d\tau$ being the proper time of the particle. From the two last formulas one can derive the equality

$$d\tau = \left(1 - \mathbf{u}^2(t)\right)^{1/2} dt, \tag{19}$$

which describes the slowing down of the course of proper time of the particle as compared to the time in the inertial frame of reference. Emphasize that the quantities $d\tau$ and $dt$ entering into (19) refer to the physically nonequivalent frames of reference: the former refers to the noninertial frame of reference $S$, and the latter to the inertial frame $K$.

Now let us turn to equality (2). The fundamental difference between (2) and (19) is that the quantities $dt'$ and $dt$ in (2) are the time intervals measured in the inertial frames of reference, which are physically equivalent to each other. It should be emphasized that by virtue of the physical equivalence of inertial frames of reference $K$ and $K'$ the primed quantities $t'$ and $\mathbf{r}'$ play in the reference frame $K'$ precisely the same role, which is played by the unprimed quantities $t$ and $\mathbf{r}$ in the reference frame $K$. Thus, the essence of the phenomenon of dynamical inhomogeneity of time described by (2) consists in that the course of time in one inertial frame of reference is changed in relation to the course of time in the other, and this change depends on the character of motion of the particle.

Turning back to the change in the course of time of a particle moving in an external electromagnetic field, it is interesting to note that for arbitrary homogeneous electromagnetic field the sum of relativistic $\mathcal{E} = mc^2$ and potential $U$ energies of particle is conserved. Indeed, from the relativistic equation of motion

$$\frac{d\mathbf{p}}{dt} = e\mathbf{E} + e[\mathbf{u}\,\mathbf{B}]$$

follows the equality $d\mathcal{E}/dt = e\mathbf{E}\mathbf{u}$. From here, by virtue of the relation $e\mathbf{E}\mathbf{u} = -(dU/dt)$, where $U = -e\mathbf{E}\mathbf{r}$ is the potential energy of particle in electric field, the required relationship is received: $d(\mathcal{E}+U)/dt = 0$, i.e. $\mathcal{E}+U = const$. As we see, the local dynamical inhomogeneity of time, which results from the motion of particle in an arbitrary homogeneous electromagnetic field, does not lead to the violation of the total energy conservation law of particle. It should be emphasized once more that the phenomenon under study is of a relative nature: the case in point is the change in the course of time of particle in one inertial frame of reference in comparison with the course of time in the other due to the motion of particle under the action of a force field.

As the physical process defining the time and distance scales we have considered in the above the motion of a point particle. Because of the fact that the velocity of particle does not exceed that of light, the quantities $dt$ and $|d\mathbf{r}|$ connected with the particle's motion make up time-like intervals: $ds^2 = dt^2 - d\mathbf{r}^2 > 0$. In the case of spatially extended particles this limitation is removed in a natural way [13,15-17]. Apparently, time in this case becomes inhomogeneous in that region of space, which is occupied by particle. If the electric charge of the particle is distributed in space according, for example, to the Gaussian distribution,

$$\rho = \rho_0 \exp\left(-\frac{|\mathbf{r}-\mathbf{r}_0(t)|}{a}\right), \qquad \rho_0 = const,$$

where $\rho = \rho(\mathbf{r},t)$ is the charge density, it should be expected that the dynamical inhomogeneity of time will be most considerable in the regions, whose linear dimensions are of the order of $a$, lying in the vicinity of maximum of the charge distribution. On the tail of the distribution the inhomogeneity will be exponentially small. The requirement for stability of particle results in that the points belonging to the region, occupied by the particle, and separated by space-like intervals cease to be physically independent. It is possible to think that this phenomenon is conditioned by the back influence of the time inhomogeneity on the physical processes occurring inside the particle. Perhaps, it is this back action on matter of the change in the course of time that results in the appearance of superluminal signals.

In the example considered above, which is illustrative of the origin of the phenomenon of dynamical change in the course of time in magnetic field, the ratio of periods of particle's motion in the reference frames $K$ and $K'$ is (see (13)): $T/T' = \gamma$. For arbitrary motion of particle the ratio of the periods of motion $T/T'$ depends not only on the velocity of relative movement of reference frames, but also on the state of motion of particle. Indeed, let in the frame $K'$ the particle move according to the law:

$$\mathbf{r}'(t') = \mathbf{a}'\sin\omega't' + \mathbf{u}'_0 t', \quad \mathbf{a}' = const, \quad \mathbf{u}'_0 = const.$$

In this case equality (2) may be written in the form:

$$dt = g(t')dt', \qquad g(t') = \gamma[1 + v_0(a'_x \omega'\cos\omega't' + u'_{0x})].$$

Calculate the period of motion $T$:

$$T = \int_0^{T'} g(t')dt' = \gamma(1 + v_0 u'_{0x})T'.$$

As we can see, the ratio $T/T'$ depends also on the velocity of motion of particle in the frame of reference $K'$.

## 3. Dynamical principle and the course of time

The conclusion that the course of time depends on the state of motion of particle, is obtained in the previous section on the basis of Lorentz transformations for spatial coordinates and time by identifying the quantities $dt$ and $|d\mathbf{r}|$ with the space and time intervals, which the point particle passes when it moves in a path. Let us show that this conclusion can be drawn mathematically more strictly - from the relativistic equations of motion and relativity principle.

In mechanics the state of particle (mass point) at an instant of time $t$ is completely determined by the position vector $\mathbf{r}$ and momentum $\mathbf{p}$, calculated at the same instant of time, i.e. by the quantities $\mathbf{r}(t)$ and $\mathbf{p}(t)$. The basic problem of dynamics is formulated as follows: given the state of motion of a particle at an instant of time $t$, one should unambiguously define its state of motion at the next instant of time $t + dt$, where $dt$ ($dt > 0$) is an infinitesimal quantity. In other words, the problem of dynamics is that the quantities $\mathbf{r}(t+dt), \mathbf{p}(t+dt)$ should be unambiguously determined by the known quantities $\mathbf{r}(t), \mathbf{p}(t)$. From the expansions into Taylor series (in the linear approximation in $dt$)

$$\mathbf{r}(t+dt) = \mathbf{r}(t) + \frac{d\mathbf{r}(t)}{dt} dt, \quad \mathbf{p}(t+dt) = \mathbf{p}(t) + \frac{d\mathbf{p}(t)}{dt} dt,$$

it follows that the basic problem of dynamics can be solved if the increment of momentum $d\mathbf{p}(t) \equiv \mathbf{p}(t+dt) - \mathbf{p}(t)$ is expressed in terms of the known quantities $\mathbf{r}(t)$ and $\mathbf{p}(t)$, i.e. if the following relationship is fulfilled

$$d\mathbf{p}(t) = \mathbf{f}(\mathbf{r}(t), \mathbf{p}(t)) dt, \qquad (20)$$

with $\mathbf{f}(\mathbf{r}(t), \mathbf{p}(t))$ being a continuous function. We suppose that the particle velocity $\mathbf{u}(t)$ can be expressed in terms of momentum $\mathbf{p}(t)$.

Let us denote by $dp(t) = (dp^0(t), d\mathbf{p}(t))$ the increment of the 4-vector of energy-momentum of a particle for the time $dt$. As there exists a relationship connecting the energy and momentum of particle, the time component of the 4-vector $dp(t)$ can be represented in the form analogous to (20). Therefore we shall use the representation

$$dp^i(t) = f^i(\mathbf{r}(t), \mathbf{p}(t)) dt, \qquad (i = 0,1,2,3), \qquad (21)$$

where $f^i(\mathbf{r}(t), \mathbf{p}(t))$ are some continuous functions. The equalities (21) express the dynamical principle of mechanics (causality principle).

By virtue of relativity principle the equalities, analogous to (21), should take place in any inertial frame of reference. Assuming the relationships (21) to hold in an inertial frame of reference $K$, we can write analogous equalities in the other inertial reference frame $K'$:

$$dp'^i(t') = f'^i(\mathbf{r}'(t'), \mathbf{p}'(t')) dt', \qquad (i = 0,1,2,3). \qquad (22)$$

Here $dp'^i(t')$ is the increment of the 4-momentum of particle for a time $dt'$ in the frame of reference $K'$ corresponding to the increment $dp^i(t)$ of the 4-momentum of the same particle for the time $dt$ in the frame of reference $K$ provided that the particle's state in the frame $K$ at the instant $t$ is described by the quantities $\mathbf{r}(t), \mathbf{p}(t)$, and in the frame $K'$ at the moment $t'$ - by the quantities $\mathbf{r}'(t'), \mathbf{p}'(t')$.

If the relation between coordinates and time in the reference frames $K$ and $K'$ is given by

$$x^i = L_{ik} x'^k, \qquad (i = 0,1,2,3). \qquad (23)$$

where $x^i = (t, \mathbf{r}) = x$, $x'^i = (t', \mathbf{r}') = x'$, $L_{ik}$ is the matrix of Lorentz transformations, the following relationships are fulfilled:

$$dp^i(t) = L_{ik} dp'^k(t'), \qquad (i = 0,1,2,3). \qquad (24)$$

The equalities (24), in which 4-vectors $dp(t)$ and $dp'(t')$ are given by (21) and (22), relate the course of time $dt$ in the frame $K$ in the vicinity of the instant $t$ at a point with position vector $\mathbf{r} = \mathbf{r}(t)$ to the course of time $dt'$ in the frame $K'$ in the vicinity of the instant $t'$ at the point with position vector $\mathbf{r}' = \mathbf{r}'(t')$ provided that the coordinates of point $x = (t, \mathbf{r}(t))$ are related to those of $x' = (t', \mathbf{r}'(t'))$ by Lorentz transformations (23). The equalities (24) by virtue of (21) and (22) can be written as

$$dt = \frac{L_{ik} f'^k(t') dt'}{f^i(t)},$$

where $f^i(t) = f^i(\mathbf{r}(t), \mathbf{p}(t))$, $f'^i(t') = f'^i(\mathbf{r}'(t'), \mathbf{p}'(t'))$. Having expressed $t$ in the right hand side of the latter equality in terms of $t'$ according to Lorentz transformations,

$$t = L_{0k} x'^k, \qquad x' = (t', \mathbf{r}'(t')) = x'^i(t'),$$

we obtain the sought-for equality:

$$dt = g_i(t') dt', \tag{25}$$

where

$$g_i(t') = \frac{L_{ik} f'^k(t')}{f^i(L_{0k} x'^k(t'))}. \tag{26}$$

To calculate the function $g_i(t')$ in explicit form, we make use of the relativistic equations of motion (see, for example, [14]):

$$\frac{dp^i}{d\tau} = e F^{ik} \frac{dx_k}{d\tau}, \tag{27}$$

where $p^i = m_0 \frac{dx^i}{d\tau}$, $d\tau = dt\sqrt{1-(\mathbf{u}(t))^2}$, $x_i = g_{ii} x^i$, $g_{00} = 1$; $g_{\alpha\alpha} = -1$ ($\alpha = 1,2,3$), $p^i$ and $m_0$ are the 4-momentum and rest-mass of particle,

$$F^{ik} = \begin{pmatrix} 0 & -E_x & -E_y & -E_z \\ E_x & 0 & -B_z & B_y \\ E_y & B_z & 0 & -B_x \\ E_z & -B_y & B_x & 0 \end{pmatrix},$$

$F^{ik}$ is the electromagnetic field tensor, $\mathbf{E}$ and $\mathbf{B}$ are electric and magnetic fields. A comparison of (27) and (21) shows that:

$$f^i(t) = e F^{ik} \frac{dx_k}{dt}, \qquad f'^i(t') = e F'^{ik} \frac{dx'_k}{dt'}. \tag{28}$$

Taking into consideration Lorentz transformations, the function $g_i(t')$ can be represented in the form

$$g_i(t) = \left( \frac{1-(\mathbf{u}'(t'))^2}{1-(\mathbf{u}(t))^2} \right)^{1/2} \Bigg|_{t = L_{0k} x'^k(t')}. \tag{29}$$

Further, for definiteness, we shall take as the reference frames $K$ and $K'$ the frames specified at the beginning of the previous chapter. By making use of the known equality (see [4])

$$\left( \frac{1-(\mathbf{u}'(t'))^2}{1-(\mathbf{u}(t))^2} \right)^{1/2} = \gamma(1 + v_0 u'_x(t')),$$

we find:

$$g_i(t') = \gamma(1 + v_0 u'_x(t')) \tag{30}$$

The equality (25), with regard to (4) and (30), results, as it should be, in relationships (2) and (3). Note that the quantity $g_i(t')$ proves to be not dependent on $i$.

It should be emphasized that the 4-vector $x = (t, \mathbf{r})$ entering into the equation of motion (27) describes a geometric point in space-time lying on the particle's trajectory. For this reason the intervals $dt$ and $|d\mathbf{r}|$, taken along the particle trajectory, have the direct physical meaning of time and space intervals relating to the physical process - the particle motion. Hence, the quantity $dt$ and the quantity $dt'$ corresponding to it in the frame $K'$ describe the course of time on the particle trajectory in the view of the $K$- and $K'$-observers. Therefore the inferences about the course of time obtained in the previous section directly from Lorentz transformations are absolutely rigorous from the physical point of view.

## 4. Conclusion

First of all, it should be noted that the proof of existence of the physical properties of time presented in this paper has not required the use of any additional hypotheses: the inferences of the paper are a direct consequence of the relativistic equations of motion and relativity principle. They can be reached immediately, if one applies the STR kinematics to the motion of a particle in space-time under the influence of a force field and takes into consideration that the time and distance scales should be associated with those time and space intervals, which really arise when the particle moves. The results of the paper are such that they could be obtained at once after the creation of the STR. Why did this not happen? The reason is the conventional Newtonian concept of a single universal time, according to which time flows absolutely independent of all material processes: "The absolute, true, and mathematical time, in itself, and from its own nature, flows equally, without relation to any thing external ... " [18]. Such a concept of time, to which the physicists were accustomed, excluded the very statement of the problem about the existence of the physical properties of time. Bowing before the genius of N. Kozyrev and rendering homage to his deep physical intuition, it should be noted that the negative attitude of the majority of physicists to his papers concerning time is explained by the fact that Kozyrev, with the aim to substantiate his views, used some additional hypotheses (for example, the hypothesis about the existence of a new constant responsible for the physical properties of time) and in doing so overstepped the limits of conventional classical mechanics. It is shown in the present paper that the idea about the existence of the physical properties of time can be proved remaining within the framework of relativistic mechanics, without introducing any additional hypotheses. The main objection of the opponents of Kozyrev's idea is thereby removed and it becomes evident that the development of research on the time as an active participant of physical processes is of fundamental importance in modern physics.

The theory given in sections 2 and 3 for point particles can be easily generalized to an arbitrary physical system. The results received allow to formulate the following general conclusion: from relativistic equations of motion, governing the behaviour of physical system, and relativity principle follows with necessity the existence of physical properties of time conditioned both by the interaction between the components of the system and by the interaction of the system with other bodies. Thus, any material system is capable to influence the course of time in that region of space, where it is placed. The ability to change the course of time during the process of motion represents one of the most fundamental properties of physical system which can be referred to as "the feeling of time". Apparently, "the feeling of time", internally inherent in any form of matter (both particles and fields) by the very nature of things underlies the specific time structure of the material world, whose existence is discussed in [6,8,9]. Let us emphasize that the existence of physical properties of time results from the pseudoeuclidity of space-time.

According to [6], the change in the course of time caused by physical processes results in the appearance of additional forces which are registered experimentally. In particular, the direction of a plumb on the earth surface is defined by the combined effect of the force of gravitational attraction, a centrifugal force of inertia, and a force associated with the change in the course of time. As is clear from the results of this paper, the forces conditioned by the physical properties of time are automatically taken into account in the relativistic equations of motion. Let us show schematically how these forces can be separated explicitly under the assumption that the velocity of relative movement of the reference frames and the particle's velocity are small as compared to the velocity of light. We proceed from the relativistic equation of motion of a particle in the reference frame $K'$:

$$\frac{d\,m\mathbf{u}'(t')}{dt'} = \mathbf{F}'(\mathbf{r}'(t'), \mathbf{u}'(t'), t').$$

Expanding both parts of this equality into a Taylor series in the vicinity of the point $t' = t$, neglecting the members of the second order of smallness, and taking into account that in the approximation considered $t' = t - v_0\,x$, we obtain the equation:

$$m_0 \frac{d\mathbf{u}'(t)}{dt} + m_0 \frac{d^2\mathbf{u}'(t)}{dt^2}(-v_0\,x'(t)) = \qquad (31)$$

$$\mathbf{F}'(\mathbf{r}'(t), \mathbf{u}'(t), t) + \frac{d\mathbf{F}'(\mathbf{r}'(t), \mathbf{u}'(t), t)}{dt}(-v_0\,x'(t)).$$

Solution to (31) can be looked for in the form

$$\mathbf{r}'(t) = \mathbf{r}_0(t) + \mathbf{r}_1(t), \qquad (32)$$

with $\mathbf{r}_1(t)$ being the correction due to the change in the course of time. The substitution of (32) into (31), after expansion into a powers series in $\mathbf{r}_1(t)$ and elementary transformations, leads to the equation (in linear approximation)

$$m_0 \frac{d^2}{dt^2}\mathbf{r}_1 = \left(\mathbf{r}_1 \vec{\nabla}_{\mathbf{r}_0}\right)\mathbf{F}'(\mathbf{r}_0, \mathbf{u}_0, t) + \left(\mathbf{u}_1 \vec{\nabla}_{\mathbf{u}_0}\right)\mathbf{F}'(\mathbf{r}_0, \mathbf{u}_0, t), \qquad (33)$$

where $\mathbf{r}_0 = \mathbf{r}_0(t)$, $\mathbf{u}_0 = \dfrac{d\mathbf{r}_0}{dt}$, $\mathbf{r}_1 = \mathbf{r}_1(t)$, $\mathbf{u}_1 = \dfrac{d\mathbf{r}_1}{dt}$. In deriving (33) it was taken into account that the quantity $\mathbf{r}_0$ obeys the following equation of zero approximation:

$$m_0 \frac{d^2}{dt^2}\mathbf{r}_0 = \mathbf{F}'(\mathbf{r}_0, \mathbf{u}_0, t).$$

The right-hand side of equality (33) gives the sought-for expression for the force. Let us note that for the construction of the consistent theory explaining the results of Kozyrev's experiments with gyroscopes and pendulums, it is necessary to fulfill an analysis of equations of motion, analogous to the one given above, for a noninertial (rotating) frame of reference.

It is pertinent here to point to an important moment regarding superluminal signals. Because the effect of the dynamical inhomogeneity of time, considered in the paper, is of a local character and the particles are point-like, the change in the course of time in the model under study cannot give rise to superluminal signals. However the latter necessarily arise when the self-action of particle is taken into account as is explained in [5,13,15-17]. The part of the physical bearer of superluminal signals is played by the own field of particle, which together with the particle makes up a single open self-organizing system. The phenomenon of dynamical inhomogeneity of time and the appearance of superluminal signals seem to represent the mutually conditioned phenomena, which can be consistently described by the non-linear quantum dynamical equation [19,20].

In the papers [13,15-17], in discussing the own field of particle, we are restricted ourselves only to the field of electromagnetic origin. It is natural to assume that there exist 4 components of the own field of material body according to four types of interaction familiar to us now - gravitational, electromagnetic, weak, and strong. Each of these components seems to be a classical (nonquantum) field, which is capable to transfer perturbation with a superluminal velocity.

In conclusion we should like to point to the applied aspect of investigation on the physical properties of time: it opens up radically new opportunities for the efficient control of physical processes.